\documentclass[preprint,showpacs,preprintnumbers,amsmath,amssymb,superscriptaddress]{revtex4}
\usepackage{amssymb}
\usepackage{graphicx,color}
\usepackage{bm}

\usepackage{graphicx} \usepackage{dcolumn} \usepackage{bm}
\newcommand{\beq}{\begin{equation}}\newcommand{\eeq}[1]{\label{#1}
\end{equation}}\newcommand{\beqar}{\begin{eqnarray}}\newcommand{\eeqar}[1]
{\label{#1}
\end{eqnarray}}\newcommand{\bmath}{\begin{displaymath}}\newcommand{\emath}{\end{displaymath}}\newcommand{\bitem}{\begin{itemize}}\newcommand{\eitem}{\end{itemize}}
 
\begin{document}

\title{\Large \bf 
Baryon anomaly and strong color fields in Pb+Pb collisions 
at 2.76A TeV at the CERN Large Hadron Collider.}

\newcommand{\columbia}{Columbia University, New York, N.Y. 10027, USA}

\newcommand{\mcgill}{McGill University, Montreal, H3A 2T8, Canada}

\affiliation{\mcgill}
\affiliation{\columbia}
\author{~V.~Topor~Pop} \affiliation{\mcgill}
\author{~M.~Gyulassy} \affiliation{\columbia}
\author{~J.~Barrette} \affiliation{\mcgill}
\author{~C.~Gale} \affiliation{\mcgill}

\date{October 14, 2011}

\begin{abstract}

With the {\small HIJING/B\=B} v2.0 heavy ion event generator, we explore the 
phenomenological consequences of several 
high parton density dynamical effects predicted in
central Pb+Pb collisions at the 
Large Hadron Collider (LHC) energies. These include (1) 
jet quenching due to parton energy loss ($dE/dx$), (2)
strangeness and hyperon enhancement due to
strong {\em longitudinal} color field (SCF), and (3)  
enhancement of baryon-to-meson ratios due to baryon-anti-baryon  junctions 
(J\=J) loops and SCF effects. The saturation/minijet cutoff scale
$p_0(s,A)$ and effective string tension $\kappa(s,A)$
are constrained by our previous analysis of LHC $p+p$ data and
recent data on the charged multiplicity for Pb+Pb collisions
reported by the ALICE collaboration. We predict the hadron 
flavor dependence (mesons and baryons) of the 
nuclear modification factor $R_{\rm AA}(p_T)$ and emphasize
the possibility that the baryon anomaly could persist at the LHC up to
$p_T\sim 10$ GeV, well beyond the range observed in central Au+Au
collisions at RHIC energies.

\end{abstract}


\maketitle


\section{Introduction}

With the commissioning of the Large Hadron Collider (LHC), it 
is now  possible to test dynamical models of multiparticle production  
in nuclear collisions up to a center-of-mass energy (c.m.) per nucleon 
$\sqrt{s_{\rm NN}}$ = 5.5 TeV.
Charged particle densities at mid-pseudorapidity, 
$(2dN_{\rm ch}/d\eta)/N_{\rm part}$,    
and their dependence on energy and centrality are important for understanding
the mechanism of hadron production and especially the interplay of soft
(string fragmentation)  and hard (perturbative quantum chromodynamics)
scattering contributions at an order of magnitude higher energies than 
were extensively studied up to now at the Relativistic Heavy Ion
Collider (RHIC/BNL).
The rate of parton-parton and multi parton-parton  
scattering and soft processes are strongly correlated with the
observed particle multiplicity (related to the
{\em initial entropy} and the {\em initial energy density} generated
in nuclear collisions over spacetime volumes up to $10^4$ 
greater than in elementary $p+p$ collisions).

Using the constraints on dynamical parameters 
from data at RHIC energies, there remain relatively large uncertainties 
($\sim$ a factor of two) on predictions for charged hadron multiplicities 
at the LHC energies, see Refs. 
\cite{Abreu:2007kv,armesto2_08,Albacete:2010fs,Kharzeev:2004if,Armesto:2004ud,Levin:2010zy,Levin:2010dw,Levin:2011hr,Deng:2010mv,Deng:2010xg,Bozek:2010wt,Sarkisyan:2010kb,Humanic:2010su,Xu:2011fi,Mitrovski:2008hb,Bopp:2007sa}.
The first data on inclusive charged particle distributions from the 
LHC in Pb+Pb collisions are now available 
\cite{Aamodt:2010pb,Aamodt:2010cz,Aamodt:2010jd,Aamodt:2010pa,Aad:2010bu,Chatrchyan:2011sx}.
Recently, ATLAS analyzed event samples from 
central (0-10\%) Pb+Pb collisions
at $\sqrt{s_{\rm NN}}$ = 2.76 TeV in which the transverse energy of
the most energetic jet (leading jet) $E_{T_1} > 100$ GeV and 
where a recoiling jet with transverse energy $E_{T_2} > 25$ GeV could 
be found at azimuthal separation $\Delta \Phi > \pi/2 $. 
In comparison with $pp$ collisions, ATLAS observed (within a narrow
cone) in Pb+Pb collisions 
a strong reduction of the number of recoiling jets carrying a large
fraction of the maximum available 
jet energy $x = E_{T_2}/E_{T_1} > 0.6 $  \cite{Aad:2010bu}. 
This is consistent with the expected strong quenching of jets
due to partonic  energy loss in ultradense quark gluon plasma (QGP).
In addition to the tomographic interest in jet quenching high $p_T$
observables, there is considerable interest on the
impact of mini-jet quenching as an additional final state 
interaction source of the observed bulk multiplicity/entropy production.
The LHC data on nucleus-nucleus ($A + A$) collisions
may lead to an improved theoretical
understanding of ultra-dense multiparton plasma
based on a quantum chromodynamics (QCD) approach 
\cite{Deng:2010mv,Deng:2010xg,Seymour:2010ih,Sassot:2010bh,CasalderreySolana:2010eh}.

To isolate quark-gluon plasma signatures in heavy-ion collision data, 
it is imperative to work in a framework that
can simultaneously account for 
nucleon-nucleon collisions ($p+p$) in the same energy range.
In fact $p+p$ reactions can not be considered as  
elementary at the LHC and their study
 has already revealed possible interesting new physics \cite{Werner:2010ss}. 
We utilize the HIJING/B\=B2.0 heavy-ion generator as a
phenomenological tool that can systematically relate observables 
in $p+p,\,\, p+A$, and $A+A$ collisions in terms
of a common framework based on a two-component picture of the dynamics
involving soft longitudinal beam jets, multiple minijets, and rare
hard pQCD processes. The Monte Carlo code inherits the phenomenological
successes of the LUND \cite{Andersson:1986gw} 
and dual parton model (DPM) \cite{Capella:1979fm} string excitation 
and hadronization
mechanisms as well as the hard processes encoded in the PYTHIA
\cite{Bengtsson:1987kr} event generator. 
Multiple low $p_T < Q_s(x,A)$ transverse momentum 
color exchanges excite the incident baryons into longitudinal strings
that fragment due to color neutralization into an array of physical 
hadron resonance states. Hard processes are modeled as high $p_T$ 
kinks in the strings and hadronize via the well-tested LUND scheme
in $e^+e^-$ and $ep$ processes. The HIJING model \cite{mik_wang_94} 
uses a variant of the string excitation
LUND and DPM models constrained to lower energy $pp$ data.

In heavy ion collisions, the novel ``nuclear physics''
is due to the nuclear modification of the parton distribution functions,
the possibility of multiple longitudinal flux tube overlapping leading to
strong longitudinal color field (SCF) effects refered to as color ropes
or glasma. Strong fields also lead to
enhanced strangeness production \cite{muller1986}.
The effect is especially enhanced for production of baryons with multiple 
strange quarks as discovered first at the SPS/CERN \cite{Andrighetto:1999gg}.
Recently, the data on the production of strange particles with one, 
two or more strange quarks for $p+\bar{p}$ collisions 
at the Tevatron \cite{Aaltonen:2011wz} 
and for $p+p$ collisions at the LHC \cite{Aamodt:2011zj,Khachatryan:2011tm} 
have been reported. In Ref.~\cite{Aaltonen:2011wz} 
it was shown that the ratio
$\Xi^-/\Lambda$ and $\Omega^-/\Lambda$ rise at low $p_T$ and the 
ratio reaches an unexpected plateau at $p_T > 4$ GeV/{\it c}, 
which persist up to rather large 10 GeV/{\it c}.
This result is a major motivation for the present work.

We address these and other issues within the framework 
of the {\small HIJING/B\=B} v2.0 model \cite{ToporPop:2010qz,prc75_top07,prc72_top05}.
A systematic comparison with data on $p+p$ and $p+\bar{p}$ 
collisions in a wide energy range \cite{ToporPop:2010qz} 
revealed that minijet production and fragmentation as implemented in the 
{\small HIJING/B\=B} v2.0 model provide a simultaneous and 
consistent explanation of several effects: the inclusive spectra at moderate 
transverse momentum ($p_T < 5$  GeV/c), the energy dependence of the central
rapidity density, the (strange)baryon-meson ratios and 
the baryon-antibaryon asymmetry.
Specifically, in this paper we extend our 
studies to the dynamic consequences of initial and 
final state saturation phenomena, gluonic J\=J loops, 
and jet quenching  
to particle production in central (0-5\%) Pb+Pb collisions at the LHC.
Our study aims to investigate a set of observables 
sensitive to the dynamics of the collisions,
covering both longitudinal and transverse degree of freedom.
We present here the pseudorapidity multiplicity density per participant
pair, and their centrality and energy dependence. 
Our study reveals a possible violation at $\sqrt{s_{\rm NN}}$ = 2.76
TeV of the  {\em limiting fragmentation} observed in 
multiparticle production of charged particles up to top RHIC energy
\cite{Busza2009,Busza:2011pc}.   
Predictions for the nuclear modification factors (NMF) of charged 
particle ($R_{\rm AA}$) and of identified particles ($R_{\rm AA}^{\rm ID}$) 
are also discussed.
We emphasize the role of SCF effects and gluonic J\=J loops in 
understanding the  {\em meson/baryon anomaly} at the LHC energy,
which manifest in the 
(strange)baryon-meson ratios at intermediate and large transverse momenta and  
in NMF of identified particles ($R_{\rm AA}^{\rm ID}$).

\section{Outline of {\small HIJING/B\=B v2.0} model.}

A detailed description of the model can be found in Refs. 
\cite{ToporPop:2010qz,prc75_top07,prc72_top05}.
The HIJING model is based on a two component geometrical 
model of minijet production and soft interaction and has incorporated 
nuclear effects such as {\em shadowing} (nuclear modification of the parton 
distribution functions) and {\em jet quenching}, via final state jet
medium interaction.  
In the {\small HIJING/B\=B} v2.0 model \cite{ToporPop:2010qz,prc75_top07}
we introduce new dynamical effects associated with
long-range coherent fields (i.e., strong longitudinal color fields, SCFs),
including baryon junctions and loops \cite{prc72_top05,ripka_lnp2004}.


Saturation physics is based on the observation
that small-{\it x} hadronic and nuclear wave functions, and thus the 
scattering cross sections as well, are described by the same internal 
momentum scale known as the {\em saturation scale} ($Q_{\rm sat}$)
\cite{larry_2009}.
A recent analysis of $pp$ data up to LHC 7 TeV \cite{McLerran:2010ex}
has shown that, with  the $k_T$ factorized gluon fusion approximation
the growth of the $dN_{ch}/d\eta$ can be accounted for
if the saturation scale grows with center-of-mass system (c.m.s.) energy as
\begin{equation}
Q_{\rm sat}^2(s) = Q_0^2 (s/s_0)^{\lambda_{\rm CGC}},
\label{eq:larry10}
\end{equation}
 with $\lambda_{\rm CGC} \approx 0.115$. 
The saturation scale is also increasing with atomic number as $A^{1/6}$
\cite{kharzeev_07} and it was argued that the effective 
string tension ($\kappa$) of color ropes should scale with
$Q_{\rm sat}^{2}$ \cite{kharzeev_07,Tanji:2008ku}. 

However, in HIJING
the string/rope fragmentation is the only soft source of multiparticle
production and multiple minijets provide a semi-hard additional
source that is computable within collinear factorized standard pQCD
with initial and final radiation (DGLAP evolution \cite{parisi_77}).  
In order to achieve a quantitative
description, within our HIJING/B\=B framework we
will show that combined effects of hard and soft sources of 
multiparticle production can reproduce the available data in the range
$0.02<\sqrt{s}< 20$ TeV only with a reduced dependence 
of the effective string tension on $\sqrt{s}$. 
We find that the data can be well reproduced taking 
\begin{equation}
\kappa(s)= \kappa_{0} \,\,(s/s_{0})^{0.06}\,\,{\rm GeV/fm}\approx
Q_{0\kappa}\,Q_{\rm sat}(s),
\label{eq:kappa_sup}
\end{equation} 
where $\kappa_{0}$ = 1 GeV/fm is the vacuum string tension value, 
$s_{0}$ = 1 GeV$^2$ is a scale factor, and 
$Q_{0\kappa}=\kappa_0/Q_0$ is adjusted
to give $\kappa = 1.88$ GeV/fm at the RHIC energy $\sqrt{s}=0.2$ TeV.
Our phenomenological $\kappa(s)$ is compared to 
$Q_{\rm sat}^2(s)$  in Fig. 1 from Ref.\cite{ToporPop:2010qz} , where 
$\kappa$ = 1.40 GeV/fm at $\sqrt{s}$ = 0.017 TeV increases to
$\kappa$ = 3.14 GeV/fm at $\sqrt{s}$ = 14 TeV.     


In a strong longitudinal color electric field, 
the heavier flavor suppression factor 
$\gamma_{Q\bar{Q}}$ varies with string tension 
via the well-known Schwinger formula \cite{schwinger}, 
\begin{equation}
\gamma_{Q\bar{Q}} = \frac{\Gamma_{Q\bar{Q}}}{\Gamma_{q\bar{q}}} =
{\text {exp}} \left(-\frac{\pi(M_{Q}^2-m_q^2)}{\kappa_0} \right)
 < 1
\label{eq:gamma_supress}
\end{equation}
for $Q = {\rm qq}$, $s$, $c$ or $b$ and $q = u$, $d$.
Therefore, the above formula implies a 
suppression of heavier quark production according to
$u$ : $d$ : ${\rm qq}$ : $s$ : $c$ $\approx$ 1 : 1 : 0.02 : 0.3 : 10$^{-11}$ 
for the vacuum string tension $\kappa_0$ = 1 GeV/fm.  
For a color rope , on the other hand,
the {\em average string tension} value increases due to 
{\em in-medium} effect.
With these increase string tensions 
the $Q\bar{Q}$ flavor pair production suppression factors, 
$\gamma_{Q\bar{Q}}$, approach unity in
$A+A$ collisions. In our model this is the main mechanism
for strange meson and hyperon enhancement.
This increase is quantified in our calculations at RHIC and LHC energies, 
using a power-law expression,  
\begin{eqnarray}
\kappa(s,A)_{\text RHIC}&=& \kappa(s) A^{0.087} \\
\kappa(s,A)_{\text LHC} &=& \kappa(s) A^{0.167}
\label{kapsA} 
\end{eqnarray}
where the exponent has been fixed to provide a good description of the
charged particle densities at mid-pseudorapidity (see below).


Moreover, the measured charged multiplicity grows faster 
($ \sim \sqrt{s}^{0.3}$) in nucleus-nucleus collisions than it does for 
protons ($\sim \sqrt{s}^{0.2}$)\cite{Collaboration:2011rt}.
The energy dependence of the multiplicity is an experimental
probe of possible {\it x} dependence of the saturation scale $Q_{\rm sat}^2$
\cite{Lappi:2011gu}. 
Also, the energy density for most central (0-5\%) collisions is larger 
by about a factor of 3 at LHC than the corresponding one at RHIC 
\cite{Collaboration:2011rt}.
Since the energy densities are computed from the square
of the field components \cite{Cardoso:2011cs} they are proportional
with $\kappa(s,A)^2$. These experimental facts indicate that we have to
use different exponent at RHIC and at LHC.
The problem was also addressed in CGC models, where 
$Q_{\rm sA}^2 \sim Q_{\rm sp}^2 (A^{1/3})^{\frac{1}{\gamma_{\rm eff}}}$.
At smaller {\it x} (corresponding to LHC energy) $1/\gamma_{\rm eff}$
is larger, and thus the nuclear enhancement of 
$Q_{\rm sat}^2$ is expected to be  
larger than at top RHIC energy \cite{Lappi:2011gu}.


It was shown recently \cite{Baier:2011ap} that charged-particle multiplicities 
measured in central heavy ion collisions at high energies may
not directly determine the initial conditions as predicted by 
CGC models. In its simplest implementation
it was generally assumed that there is a direct correspondence
between the number of partons in the initial state 
and the number of particles in the final state 
\cite{larry_2009,McLerran:2010ex}. However, final-state nonequilibrium 
production mechanisms generally increase the final multiplicity.
In fact, in recent HIJING2.0 (which includes modern structure
functions but without SCF and J\=J loops) the enhancement 
of the multiplicity due to quenching of final minijets had to be
turned off not to over predict the ALICE $dN_{ch}/dy$ \cite{Deng:2010mv}.
In HIJING/B\=B2.0 used here with supersaturated Duke-Owens 
parton distribution functions we 
find a consistent description of both $p+p$ and Pb+Pb multiplicity when
taking into account the extra multiplicity generated by minijet quenching.

As mentioned, in HIJING the string/rope fragmentation is the only 
{\em soft source} of multiparticle
production and multiple minijets provide a semihard additional
source that is computable within collinear factorized standard pQCD
with initial and final radiation (DGLAP evolution \cite{parisi_77}).  
Within the HIJING model, one assumes that nucleon-nucleon
collisions at high energy can be divided into {\em soft} and
{\em hard} processes with at least one pair of jet production with
transverse momentum, $p_{T}>p_0 $. A cutoff (or saturation) 
scale $p_0(s,A)$ in the final jet production has to be introduced 
below which the high density of initial interactions
leads to a nonperturbative mechanisms which in the HIJING framework
is  characterized by
a finite soft parton cross section $\sigma_{\rm soft}$.
The inclusive jet cross section $\sigma_{\rm jet}$ at leading order
(LO) \cite{Eichten:1984eu} is 

\begin{equation}
\label{eq:sigma_jet}
 \sigma_{jet}=\int_{p_0^2}^{s/4}\mathrm{d}p_T^2\mathrm{d}y_1\mathrm{d}y_2 
 \frac{1}{2}\frac{\mathrm{d}\sigma_{jet}}{\mathrm{d}p_T^2
 \mathrm{d}y_1 \mathrm{d}y_2},
\end{equation}
where,
\begin{equation}
 \frac{\mathrm{d}\sigma_{jet}} {\mathrm{d}p_T^2 \mathrm{d}y_1 
\mathrm{d}y_2} =K \sum_{a,b}x_1f_a(x_1,p_T^2)x_2f_b(x_2,p_T^2) 
\frac{\mathrm{d}\sigma^{ab}(\hat{s},\hat{t},\hat{u})}{\mathrm{d}\hat{t}}
\end{equation}

depends on the parton-parton cross section $\sigma^{ab}$ and parton 
distribution functions (PDF), $f_a(x,p_T^2)$. The summation runs over all 
parton species; $y_1$ and $y_2$ are the rapidities of the scattered
partons; $x_1$ and $x_2$ are the light-cone momentum 
fractions carried by the initial partons.
The multiplicative $K$ factor ($K \approx 1.5-2$) accounts for    
the next-to-leading order (NLO) corrections to the leading order (LO)
jet cross section \cite{Eskola:1995cj,Campbell:2006wx}. 
The parton distributions per nucleon in a nucleus (with atomic number
$A$ and 
charge number $Z$), $f_{a/A}(x,Q^2,r)$, are assumed to be
factorizable into parton distributions in a nucleon ( $f_{N/A}$)
and the parton shadowing factor ($S_{a/A}$),

\begin{equation} 
f_{a/A}(x,Q^2,r) = S_{a/A}(x,r)f_{N/A}
\end{equation}

For the shadowing factor  $S_{a/A}$ we take the parametrization used 
in the regular HIJING model \cite{mik_wang_94,Wang:1991xy} . 
We also take into account 
the intrinsic transverse momentum and the transverse momentum broadening 
due to initial multiple scattering.

A saturation in the final state, i.e., of produced gluons,
is possible, even without requirement of saturation in the initial
state \cite{eskola_2002}. 
The largest uncertainty in minijet cross sections is the strong 
dependence on the minimum transverse momentum scale cutoff, $p_0$.   
It was shown that an increased value (with c.m. energy $ \sqrt{s} $)
is required by the experimental data indicating that 
the {\em coherent interaction} becomes important. This might be taken also 
as an evidence of gluon final state saturation at very small {\it x} 
\cite{Deng:2010mv,Deng:2010xg}. 
With the Duke-Owens parametrization \cite{Duke:1983gd} 
of parton distribution functions, an energy-independent
cutoff scale $p_0$ = 2 GeV/{\it c} and a constant soft
parton cross section $\sigma_{soft}$ = 57 mb are found to
reproduce the experimental data on 
the hadron central rapidity density in $p+p(\bar{p})$ collisions 
\cite{ToporPop:2010qz}.


However, with a constant momentum cutoff $p_0$ = 2 GeV/{\it c}, 
the total number of minijets per unit transverse area 
for independent multiple jet production in central nucleus-nucleus
collisions, could exceed the limit \cite{Deng:2010mv,Deng:2010xg}
\begin{equation}
\frac{T_{AA}(b) \sigma_{\rm jet}}{\pi R_{A}^{2}} \leq 
\frac{p_0^2}{\pi}
\end{equation}  
where $T_{AA}(b)$ is the overlap function of
$A+A$ collisions and $\pi/p_0^2$ is the intrinsic 
transverse size of a minijet with transverse momentum $p_0$.
Therefore, we consider an energy and nuclear size dependent 
cutoff $p_0(s,A)$, to ensure the applicability
of the two-component model in the HIJING model for $A+A$ collisions.
The pseudorapidity distribution of charged particle in central 
nucleus-nucleus collisions at RHIC and LHC 
energies can well be fitted (see Sec. IIIA) 
if we consider a scaling law of the type
$C\,A^{\alpha}\,\sqrt{s}^{\beta}$:

\begin{equation}
p_0(s,A)= 0.416\,A^{0.128}\,\sqrt{s}^{\,\,0.191}\,\,{\rm GeV}/c
\label{p0}
\end{equation}

New values for $p_0(s,A)$ increase from $p_0$ = 1.5 GeV/{\it c} at  
the c.m.s. energy $\sqrt{s}$ = 0.02 TeV up to $p_0$ = 4.2 GeV/{\it c} 
at $\sqrt{s}$ = 5.5 TeV.
These values obtained for central $A+A$ collisions are not expected  
to be valid for peripheral $A+A$ or $p+p$ collisions.
These dependences are similar to those used
in pQCD+saturation model from Ref.~\cite{Eskola:1999fc}. 
The main difference is a factor of two in the constant C, 
$C_{\rm HIJ}= 0.416$ in comparison with $C_{esk}=0.208$ 
from Ref.~\cite{Eskola:1999fc}.
Note that $C_{esk}=0.208$ result in an overestimate of the
charged particle density at LHC energies by a factor of approximately 
two \cite{Eskola:1999fc}.

The above limit for incoherent minijet production should also depend
on impact parameter (b) \cite{Eskola:2000xq}. 
In the present calculations within our model such dependence is not 
included.
Instead, in the  HIJING model an impact parameter dependence 
of the gluon shadowing is considered in the parametrization of the 
parton shadowing factor $S_{a/A}$ \cite{Wang:1991xy}.  
Due to shadowing effects the observed $A$ exponent ($\alpha=0.128$) 
is somewhat less than the 
number expected in the saturated scaling limit ($p_0(s,A) \sim A^{1/6}$)
\cite{Eskola:1999fc}.



The ALICE experiment at the LHC published first 
experimental data on the charged 
hadron multiplicity density at mid-rapidity in central (0-5\%) Pb + Pb 
collisions at $\sqrt{s_{\rm NN}}$=2.76 TeV \cite{Aamodt:2010pb}.
In this experiment the collaboration also confirmed the presence of 
jet quenching \cite{Aamodt:2010jd} by studying $R_{AA}$.  
A measure of the jet quenching at large transverse momentum ($p_T$)
is given by the ratio of particle yield in $A+A$ collisions
to that in $p+p$ collisions, defined by
\begin{equation}
\label{eq:RAA}
R_{AA}\,=\,\frac{(1/N_{\rm evt}^{AA})d^{2}\,N_{AA}/d^{2}\,p_{T}dy}
           {N_{\rm coll}(1/N_{\rm evt}^{pp})d^{2}\,N_{pp}/d^{2}\,p_{T}dy}
\end{equation} 

where, $N_{\rm evt}$ is the number of events and $N_{\rm coll}$
is the number of binary nucleon-nucleon ($NN$) collisions.
These results provide stringent 
constraints on the theoretical uncertainties in the bulk hadron production
and have important consequences on theoretical predictions for
jet quenching in Pb + Pb collisions at LHC energies.

For a parton $a$, the energy loss per unit distance can be expressed as   
$dE_a/dx = \epsilon_a/\lambda_a$,
where $\epsilon_a$ is the radiative energy loss per scattering and 
$\lambda_a$ the mean free path (mfp) of the inelastic scattering.
For a {\em quark jet} at the top RHIC energy
$(dE_q/dx)_{\rm RHIC} \approx 1$ GeV/fm and mfp 
$(\lambda_q)_{\rm RHIC} \approx 2$ fm 
\cite{prc75_top07}. The initial parton density is 
proportional to the final hadron multiplicity density.
The charged particle density at mid-pseudorapidity at the LHC is
a factor of 2.2 higher than that at the top RHIC energy \cite{Aamodt:2010pb}. 
Therefore, for a {\em quark jet} at the LHC  
the energy loss (mfp) should increase
(decrease) by a factor of $\approx 2.0$ and become 
$(dE_q/dx)_{\rm LHC} \approx 2$ GeV/fm and mfp 
$(\lambda_q)_{\rm LHC} \approx 1$ fm.
Throughout this analysis we will consider the results 
with two sets of parameters for hard interactions:\\ 
$(dE/dx)_1$, i.e., $K = 1.5$; $dE_q/dx = 1$ GeV/fm; $\lambda_q=2 $ fm
and \\
$(dE/dx)_2$, i.e., $K = 1.5$; $dE_q/dx = 2$ GeV/fm; $\lambda_q=1 $ fm.

Since there is always a coronal region with an average length of $\lambda_q$
in the system where the produced parton jets will escape without scattering or 
energy loss, the suppression factor can never be infinitely small.
For the same reason, the suppression factor also depends on $\lambda_q$.
It is difficult to extract information on both $dE_q/dx$ and $\lambda_q$
simultaneously from the measured spectra in a model independent way
\cite{Wang:1998bha}.
 We show below that a constant radiative energy loss mechanism 
(dE/dx=const) as 
implemented in {\small HIJING/B\=B v2.0} model provides a good description of 
suppression at intermediate and larger $p_T$ ($4< p_T < 20$ GeV/{\it c})
for charged particle in Pb+Pb collisions at the LHC.

\section{Results and Discussion}

\subsection{Multiplicity and centrality dependence}

Charged-particle multiplicity density ($2dN_{\rm ch}/d\eta)/N_{\rm part}$)  
in central (0-5\%) Pb+Pb collisions 
at $\sqrt{s_{\rm NN}}$=2.76 TeV reported by the ALICE Collaboration
is a factor of 2.2 higher than that 
observed in central Au+Au collisions at top RHIC energy.
This value is larger than most of theoretical predictions,
especially those from the CGC models \cite{Aamodt:2010pb}.
Such large hadron multiplicity will have important consequences 
on the estimated {\em shadowing} and {\em jet quenching} in Pb+Pb collisions 
\cite{Che:2011vt}. 
The value of $(2dN_{\rm ch}/d\eta)/N_{\rm part}$ at mid-pseudorapidity was
reproduced by the {\small HIJING}v2.0 model with a new set of PDFs 
\cite{Deng:2010mv} and has helped to constrain the gluon shadowing parameter
in the model \cite{Deng:2010xg}. However, these calculations do not 
include the effects of {\em jet quenching}, which has been reported 
recently by ALICE \cite{Aamodt:2010jd}, ATLAS \cite{Aad:2010bu}   
and CMS \cite{Chatrchyan:2011sx}.

\begin{figure} [h!]

\centering

\includegraphics[width=0.7\linewidth,height=9.0cm]{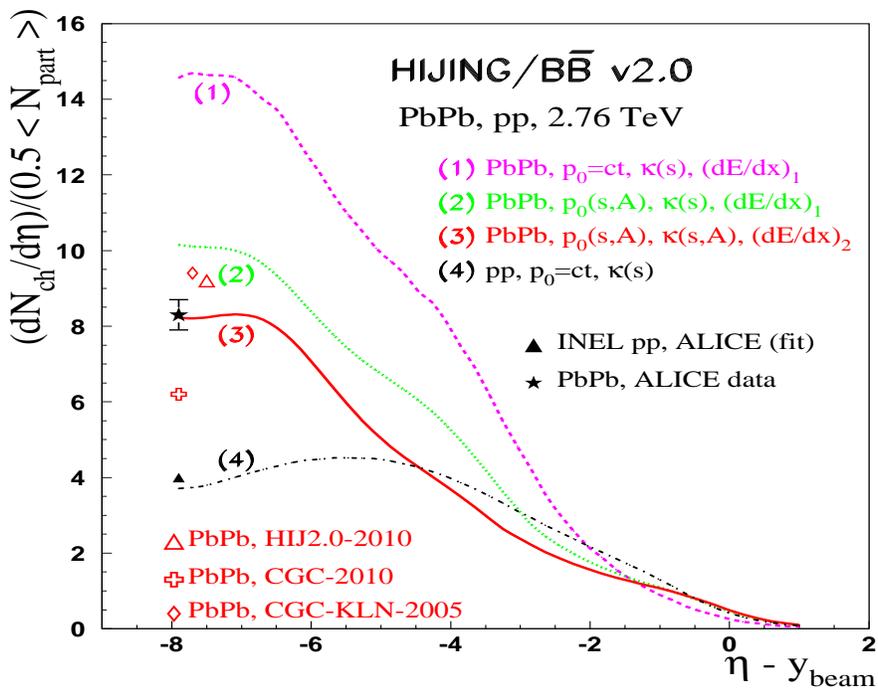}
\vskip 0.5cm\caption[pseudorapidity density per part at 2.76 TeV]
{\small (Color online) Comparison of {\small HIJING/B\=B v2.0} 
predictions for $(2dN_{\rm ch})/d\eta)/N_{\rm part}$ in 
central (0-5\%) Pb+Pb collisions at $\sqrt{s_{\rm NN}}$ = 2.76 TeV. 
The results which give the best fit of the 
experimental value \cite{Aamodt:2010pb} are plotted as a solid curve.
The dotted and dashed curves illustrate the sensitivity to the 
main model parameters (see Sec. IIIA for explanation).
The dot-dashed curve is obtained for $p+p$ NSD collisions  
at $\sqrt{s_{\rm NN}}$ = 2.76 TeV.
The full triangle is the result obtained by interpolating data between
2.36 and 7 TeV in Ref. \cite{Aamodt:2010cz}.
Comparison with other model calculations are also shown: 
{\small HIJING} v2.0 model (open triangle from Ref. \cite{Deng:2010xg}),
CGC models (open cross from Ref. \cite{Levin:2010zy}; open diamond
from Ref. \cite{Kharzeev:2004if}). 
\label{fig:fig1}
}

\end{figure}

In previous studies of heavy-ion collisions at RHIC 
\cite{ToporPop:2010qz,prc75_top07}, 
the parameters of the {\small HIJING/B\=B} v2.0 model
were adjusted in order to reproduce the measured charged particle multiplicity
as well as (multi)strange particle production.
In the present study of heavy-ion collisions at the LHC we have to
modify the values of these parameters in order to fit ALICE data 
\cite{Aamodt:2010pb}.
As shown in Fig.~\ref{fig:fig1} a good description (solid curve) 
of the measured charged particle multiplicity density at mid-pseudorapidity  
is obtained if the parameters $p_0(s,A)=3.7$ GeV/{\it c}, 
$\kappa(s,A)$ = 6.0 GeV/fm, and 
$(dE/dx)_2$ (i.e., $K=1.5$; $dE_q/dx=2$ GeV/fm; $\lambda_q=1.0$ fm) are used.

The sensitivity of the calculations to the main parameters is illustrated 
by the dashed, dotted, and solid curves. The dashed curve is obtained with 
the parameters used at RHIC energies: 
$p_0=2.0$ GeV/{\it c}, $\kappa(s)$ = 2.6 GeV/fm
and $(dE/dx)_{1}$ (i.e., $K=1.5$; $dE_q/dx=1$ GeV/fm; $\lambda_q=2.0$ fm).
The dotted curve is obtained if $p_0$ is increased 
to $p_0(s,A)=3.7$ GeV/{\it c}.
Changing from set $(dE/dx)_1$ to $(dE/dx)_2$ results 
in a good description of data (solid curve).
The dot-dashed curve is the results for $p+p$ non-single diffractive 
(NSD) collisions at $\sqrt{s_{NN}}$=2.76 TeV and are obtained with  
the same set of parameters as described in Ref.~\cite{ToporPop:2010qz}.  

The study of the dependence of observable $(2dN_{\rm ch}/d\eta)/N_{\rm part}$
on the colliding system, center-of-mass energy, and collision geometry is 
important to understand the relative contributions to particle production of 
hard scattering and soft processes.
Multiparticle production of charged particles at RHIC energies exhibit 
the phenomenon of {\em limiting fragmentation} \cite{Busza2009,Busza:2011pc}.
This phenomenon is a consequence of Feynman scaling
(if one views the collision in the rest frame of one of the incident 
particles, the production process of the soft particles is independent 
of the energy or rapidity of the other particle.). 
The region in rapidity over which the particles appears to be independent
of energy, increases with energy.
It was argued that this phenomenon 
(called {\em extended longitudinal scaling}),
is a direct manifestation of initial saturation which is assumed 
in CGC models \cite{Busza:2011pc}.

\begin{figure}[h!]
\centering
\includegraphics[width=0.8\linewidth,height=10.0cm]{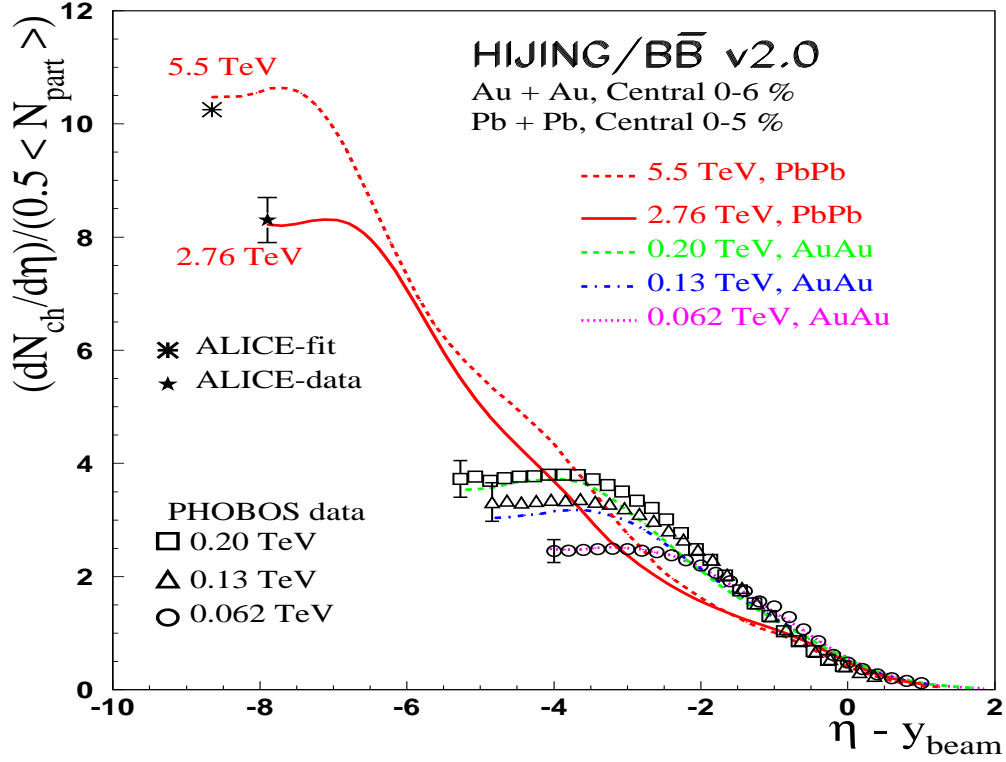}
\vskip 0.5cm\caption[Central particle density-energy dependence]
{\small (Color online) Comparison of {\small HIJING/B\=B v2.0} 
predictions for central charged particle density 
per participant pair produced in $A+A$ collisions at various energies.
The data and theoretical results are plotted in the rest frame of one of 
the nuclei.
The Au+Au data are from the PHOBOS Collaboration \cite{Busza2009,Busza:2011pc}.
The statistical error bars are plotted only at mid-pseudorapidity for clarity.
The Pb+Pb data at  $\sqrt{s_{\rm NN}}$=2.76 TeV (filled star)   
are from the ALICE Collaboration \cite{Aamodt:2010pb}.
The value at $\sqrt{s_{\rm NN}}$= 5.5 TeV (star) is obtained by a power-law 
extrapolation from RHIC energies and 2.76 TeV data \cite{Aamodt:2010pb}.
\label{fig:fig2}
}
\end{figure}

Figure 2 display the results for central (0-6\%) Au+Au collisions 
at RHIC energies (0.062 TeV $< \sqrt{s_{\rm NN}} < 0.20 $ TeV) and 
in central (0-5\%) Pb+Pb collisions at  
LHC energies ($\sqrt{s_{\rm NN}}$=2.76 TeV and $\sqrt{s_{\rm NN}}$=5.5 TeV). 
Our calculations show that at RHIC energies the model 
predicts approximately a scaling seen also in data.
However, some degree of violation of the phenomenon of 
{\em limiting fragmentation} and of the 
{\em extended longitudinal scaling} is predicted at both LHC energies.
This violation is due to the parton hard scattering included in our model
and missing in CGC models 
\cite{McLerran:2010ex,Levin:2010zy,Albacete:2010fs,Kharzeev:2004if}.

\begin{figure}[h!]
\centering
\includegraphics[width=0.8\linewidth,height=10.0cm]{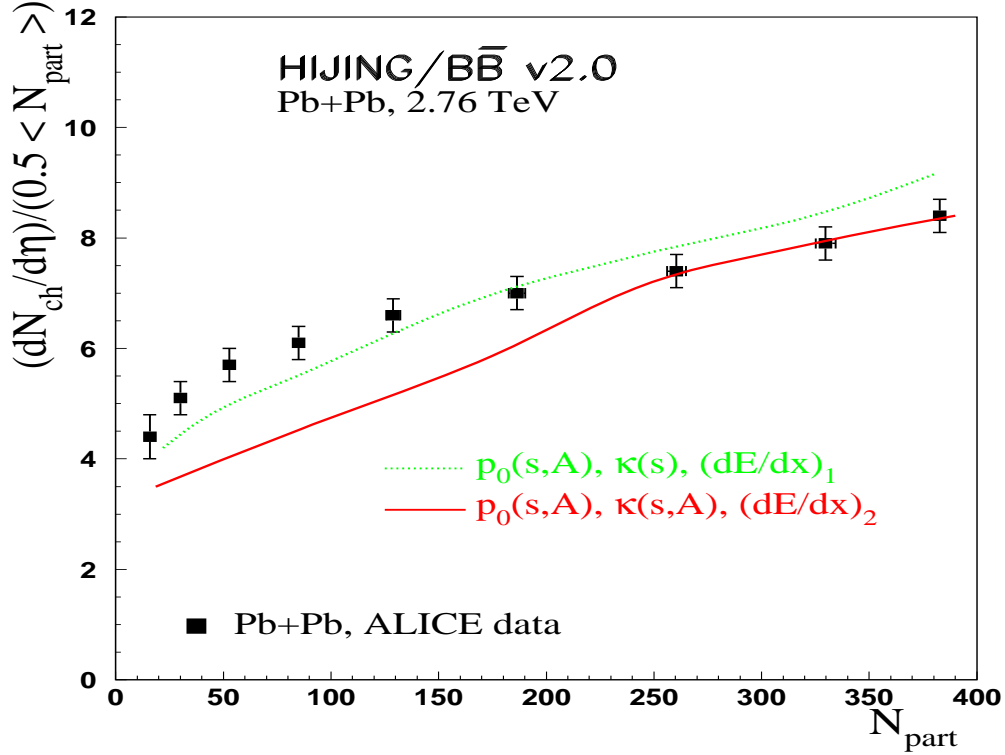}
\vskip 0.5cm\caption[Centrality dependence of charged particle density]
{\small (Color online) Comparison of {\small HIJING/B\=B v2.0} 
model calculations for charged particle multiplicity density
at mid-pseudorapidity per participant pair   
$(2dN_{\rm ch}/d\eta)/N_{\rm part}$ as a function of number of participants  
in Pb+Pb collisions at $\sqrt{s_{\rm NN}}$ = 2.76 TeV.
The solid and dotted  curves are the results obtained using 
the same set of parameters as 
in Fig.~\ref{fig:fig1} (see text for explanations).
All calculations include SCF effects and J\=J loops.
The data are from ALICE Collaboration \cite{Aamodt:2010cz} and 
the error bars include statistical and systematic uncertainties. 
\label{fig:fig3}
}
\end{figure}

Figure 3 shows the dependence of the pseudorapidity densities 
($(2dN_{\rm ch}/d\eta)/N_{\rm part}$) at mid-pseudorapidity, 
on the number of participants as measured by the 
ALICE Collaboration. 
The charged-particle density normalized per participant pair increases by
a factor of roughly two from peripheral (70-80\%) to central (0-5\%)
Pb+Pb collisions.  
The model calculations which have been tuned to high-energy $p+p$ 
\cite{ToporPop:2010qz} and central (0-5\%) Pb+Pb collisions data reasonably 
describe the central Pb+Pb data for $N_{\rm part} > 250$.
However, with the same set of parameters the model predicts 
a stronger rise with centrality than observed.

For peripheral collisions a better description is obtained 
if one use as parameters for the soft part 
$\kappa(s)$ (instead of $\kappa(s,A)$), successful in describing $p+p$ 
collisions, and for hard scattering $p_0(s,A)$, $(dE/dx)_{1}$,
as deduced from RHIC data.  
However, in this scenario the model overpredicts the absolute magnitude
for central collisions ($N_{\rm part} > 250$). These results indicate that
the average energy loss per unit distance and the effective
string tension values should have an impact parameter ($b$) 
and a parton density dependence, and a fine-tuning is required to 
describe better these data. We leave this study to a future analysis.

\subsection{Nuclear modification factor for charged particle}

In this paper, we make a simultaneous analysis of the high and low $p_T$ part 
of charged-hadron $p_T$ spectrum. 
Experiments at RHIC energies reported a suppression by a factor 
$\approx$ 5 compared 
with expectations from an independent superposition of nucleon-nucleon
($NN$) collisions \cite{Adams:2005dq,Adcox:2004mh}, interpreted as  
strong evidence of formation of a strong-coupling QGP (sQGP) 
in $A+A$ collisions.
The jet quenching has also been confirmed in central (0-5\%) 
$Pb+Pb$ collisions at the LHC \cite{Aamodt:2010jd,Aad:2010bu,Chatrchyan:2011sx}.

Having a good description for charged particle multiplicity 
at mid-pseudorapidity,
we analyze the model predictions on suppression of single hadron spectra 
in central (0-5\%) Pb+Pb collisions at the LHC.
For other recent works discussing NMF of charged particle ($R_{\rm AA}$)
at the LHC, see recent references   
~\cite{Che:2011vt,Majumder:2011uk,Renk:2011gj,Kormilitzin:2010kr,Srivastava:2011nq,Horowitz:2011gd,Levai:2011qm}.  
At the higher LHC energies a larger density of the medium is expected. 
This should lead to a larger suppression compared with that seen 
at RHIC energies.
The observed nuclear modification factor $R_{\rm AA}$ \cite{Aamodt:2010jd}  
is characterized by the following behavior:
a maximum at approximately $p_T = 2 $ GeV/{\it c} and a decrease 
with increasing $p_T$ in the range 2-6 GeV/{\it c}.
For $p_T$ greater that 6 GeV/{\it c}, $R_{\rm AA}$ rises to reach a value of 
$\approx 0.4$ at $p_T \approx 20$ GeV/{\it c}.

\begin{figure} [h!]

\centering

\includegraphics[width=0.9\linewidth,height=7.5cm]{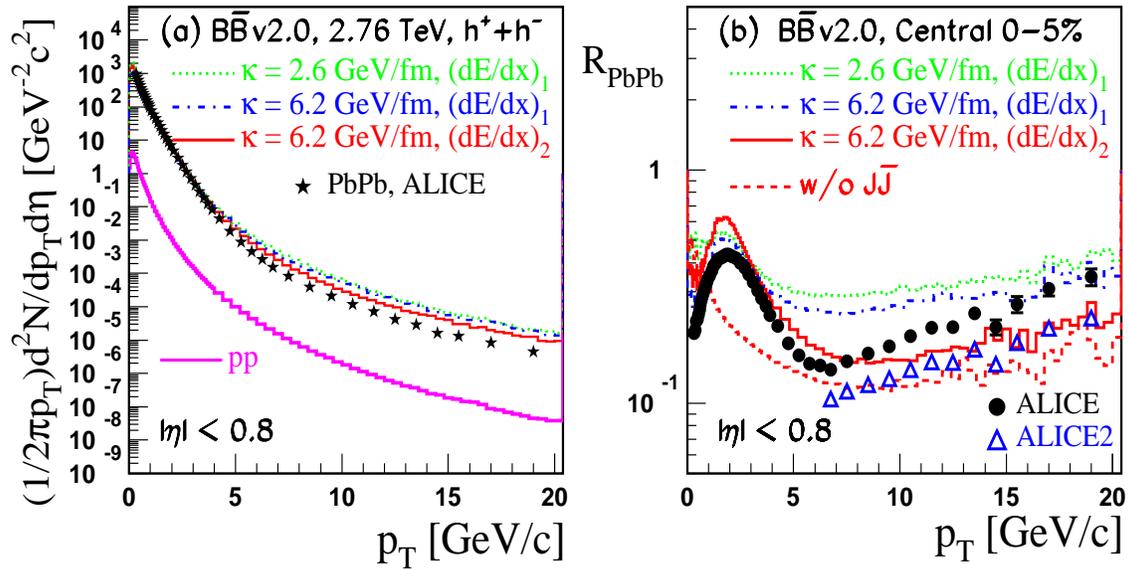}
\vskip 0.5cm\caption[R_AA for primary charged particles Pb_Pb at 2.76 TeV]
{\small (Color online) 
 (a) Comparison of {\small HIJING/B\=B v2.0} predictions for
the $p_T$ distributions of primary charged particles at mid-pseudorapidity
($-0.8<\eta<0.8$) in central (0-5\%) Pb+Pb collisions at 
$\sqrt{s_{\rm NN}}$ = 2.76 TeV.
Theoretical calculations are for different values of parameters (see text for
explanations). The $p+p$ references values are shown as the 
lower solid histogram.  
 (b) Theoretical calculations for nuclear modification factors $R_{\rm AA}$
are compared to data.
The data (left and right panels) are from the ALICE Collaboration 
\cite{Aamodt:2010jd}.
Error bars are statistical only . In part (b) the open triangles 
are the results for $R_{\rm AA}$ at $p_T > 6.5$ GeV/{\it c} using 
alternative $p+p$ reference spectrum \cite{Aamodt:2010jd}.
\label{fig:fig4}
}

\end{figure}

The measured and predicted 
transverse momentum spectra at mid-pseudorapidity 
of primary charged particles in
central (0-5\%) Pb+Pb collisions (upper solid histogram) and 
in $p+p$ collisions (lower solid histogram) at 
$\sqrt{s_{\rm NN}}$=2.76 TeV are compared in Fig.~\ref{fig:fig4}(a).
The {\small HIJING/B\=B} v2.0 model describe fairly 
well the experimental $p_T$ spectrum reported in Ref.~\cite{Aamodt:2010jd}. 
The sensitivity of the calculations to the 
effective value of the string tension ($\kappa$) is shown by
comparing the dotted ($\kappa(s) =2.6 $ GeV/fm) and the
dashed ( $\kappa(s,A) =6.2 $ GeV/fm)  histograms.  
The sensitivity to the main hard scattering parameters 
is seen by comparing the 
results obtained with the two sets $(dE/dx)_2$ (solid histograms) and  
$(dE/dx)_1$ ( dash-dotted histograms).
However, with the best set of parameters (solid histograms) 
the model still overestimate the high $p_T$ tail of the $p_T$ spectra.
There are no experimental data on charged hadron spectra in $p+p$ 
collisions at $\sqrt{s_{\rm NN}}$ = 2.76 TeV. ALICE data on the suppression 
factor ($R_{AA}$) were obtained  using a $p+p$ spectrum interpolated from 
their data at 0.9 TeV and 7 TeV.  
The open triangles in Fig.~\ref{fig:fig4}(b) 
represent the uncertainty in the interpolation method (see
Ref.~\cite{Aamodt:2010jd} for more details).

In medium modification of NMFs $R_{\rm AA}$ is mainly caused by 
the energy loss suffered by partons while transversing the plasma due 
to collisions and radiation of gluons before they fragment 
and by the nuclear shadowing. 
To describe the nuclear shadowing we use in Eq.~(8) the 
same functions as those used in  regular HIJING model \cite{mik_wang_94}.
This has been shown to be successful in explaining the results for  
$R_{\rm AA}$ at RHIC energies \cite{prc72_top05,prc75_top07}.
The benchmark of any theoretical calculation is to describe the shape seen 
at low $p_T < 6 $ GeV/{\it c} in $R_{\rm AA}$.
We can explain the shape and the presence of a maximum 
at approximately $p_T = 2 $ GeV/{\it c}   
as a specific interplay between  nonperturbative (SCF effects) 
and perturbative (gluonic J\=J dynamical loops) mechanism contributions. 
The results without J\=J loops (dashed histogram)
shown in Fig.~\ref{fig:fig4}(b) strongly underestimate the experimental values 
of $R_{\rm AA}$ in this $p_T$ range.

Since we are mostly interested in the overall effects, we concentrate 
also on the modification of high $p_T$ hadron spectra
due to an assumed total energy loss 
related to the averaged energy loss per unit distance. 
The mechanism which gives a constant energy loss per collisions is seen to
describe quite well the data (within the systematics uncertainties) 
over the largest $p_T$ window of 6 - 20 GeV/{\it c}. 
$R_{\rm AA}$ increases with $p_T$ because of the constant energy loss 
assumed here.
For a constant energy loss per collisions ($\Delta E_T$) 
the ratio $\Delta E_T/E_T$ becomes smaller for larger $E_T$, and thus 
the suppression is expected to decrease with increasing $p_T$.
Here, the energy loss per collisions is about twice as large  
and the mean-freepath of parton-parton interactions 
is smaller by a factor of roughly
two as compared with that used at the top RHIC energy.
These analyses can provide information about 
the average total energy loss the parton suffers during 
its interaction with the medium. However, a more quantitative analysis 
should be performed with the knowledge of the dynamical evolution
of the system which is beyond the scope of this paper.

\subsection{Baryon/meson anomaly at the LHC}

Identified particles at high $p_T$ 
provide direct sensitivity to differences 
between quark and gluon fragmentation \cite{Abelev:2006jr}.
Proton and pion production at high $p_T$ is expected to have significant 
contributions from quark 
fragmentation while antiprotons result mainly from gluon fragmentation
\cite{Wang:1998bha}. Therefore, the ratios $\bar{p}/\pi^-, \,\,p/\pi^+$
are sensitive to the possible color charge dependence of energy loss.
Systematic measurements of identified particle spectra in $p+p$ and Au+Au
collisions at RHIC energies 
\cite{Abelev:2006jr,:2008ez,Abelev:2007ra,Adler:2003kg,Adams:2003am} 
in the $p_T$ region 2-12 GeV/{\it c} show a significant suppression 
with respect 
to binary scaling. However, protons and anti-protons are less suppressed than 
pions at intermediate $p_T$ (2-9 GeV/{\it c}). For $p_T>2$ GeV/{\it  c}, 
the ratios $\bar{p}/\pi^-$ and  $\,\,p/\pi^+$ 
do not depend strongly on energy and they are higher in central than in 
peripheral collisions. 
In addition to being sensitive to quark and gluon jet production these 
are also sensitive to baryon transport properties and the energy density.

Moreover, the meson/baryon ratio suggest that 
baryon and antibaryon production may dominate the moderate high $p_T$ flavor
yields. These findings challenge various models incorporating jet quenching
\cite{Wang:1998bha},\cite{Vitev:2001zn},  
and/or constituent quark coalescence \cite{Fries:2003kq,Fries:2008hs}.
The fact that in the intermediate $p_T$ region 
the $p/\pi^+$ and $\bar{p}/\pi^-$
ratios are close to unity ( baryon/meson anomaly) has been attributed to 
either quark coalescence 
or to novel baryon transport dynamics \cite{Vitev:2001zn}, based on
topological gluon field configurations 
\cite{Rossi:1977cy,Montanet:1980te,Kharzeev:1996sq,Vance:1999pr}.
However, the baryon junction model predictions 
from Ref.~\cite{Vitev:2001zn} are not 
in agreement with data \cite{Abelev:2007ra}.

\begin{figure} [h!]

\centering

\includegraphics[width=0.9\linewidth,height=7.5cm]{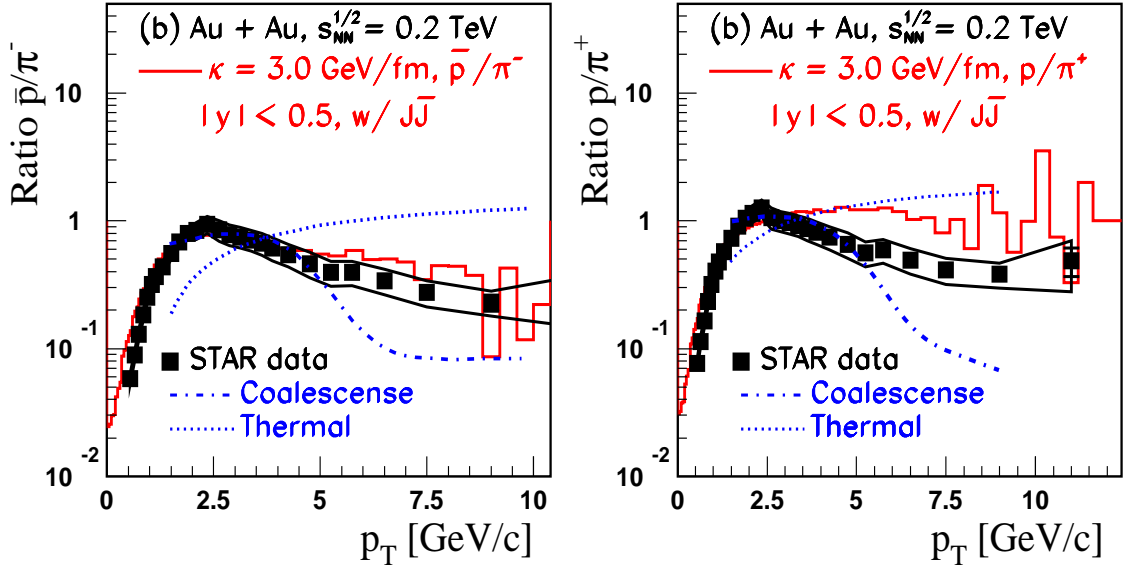}
\vskip 0.5cm\caption[ratio pbar/pi-, p/pi+]
{\small (Color online) Comparison of {\small HIJING/B\=B v2.0} 
predictions (solid histogram) for ratios $\bar{p}/\pi^{-}$ (left panel) and 
$p/\pi^{+}$ (right panel) to STAR data \cite{Abelev:2007ra} and to coalescence 
(dot-dashed curves)
 and thermal (dotted curves) model calculations from Ref.~\cite{Fries:2003kq}.
The error bars include only statistical uncertainties and systematics 
are introduced by thin continuous curves.
\label{fig:fig5}
}

\end{figure}

Figure~\ref{fig:fig5} displays the mid-rapidity ratios for central 
(0-12\%) Au+Au collisions \cite{Abelev:2007ra} at 
$\sqrt{s_{\rm NN}}$ = 0.2 TeV in comparison with model predictions.
The observed ratios peak at $p_T \approx 2 - 3 $ GeV/{\it c} with 
value close to unity and then  
decrease with increasing $p_T$ and approach the ratios observed in 
$p+p$  collisions at $p_T > 6$ GeV/{\it c}.
The results presented (solid histograms) include  
SCF and J\=J loops contributions.
The calculations based on quark coalescence (dot-dashed curves) and 
thermal models (dotted curves) from Ref.~\cite{Fries:2003kq} are
also plotted.
The grouping of particle production according to the number of constituent 
quarks has been attributed to quark coalescence from a collective 
partonic medium \cite{Fries:2008hs}.
The coalescence models \cite{Fries:2003kq,Fries:2008hs}  
can qualitatively describe these data at intermediate $p_T < 5$
GeV/{\it c}, but strongly underpredict the results at high 
$ p_T $, due to fragmentation functions used within the model.
To the contrary,  
the thermal model calculations \cite{Fries:2003kq} strongly
overestimate the data at high $p_T$. 
Our model provides an alternative explanation to the baryon/meson anomaly.
The reason is a specific interplay of 
contributions from SCF effects and dynamical J\=J loops. 
The predictions of our model are in good agreement with data 
up to the highest $p_T$, in the limit of systematics uncertainties. 
However, the model overestimate the $p/\pi^+$ ratio 
for $3 < p_T < 6$ GeV/{\it c}, while it provides a good description for
protons ($p$), it underpredicts the pion ($\pi^+$) spectrum  in this region.
Since the ratio $\bar{p}/\pi^-$ is well described by the model, 
these calculations could indicate that the fragmentation functions 
for $\pi^+$ production are not well estimated.

\begin{figure} [h!]

\centering

\includegraphics[width=0.9\linewidth,height=7.5cm]{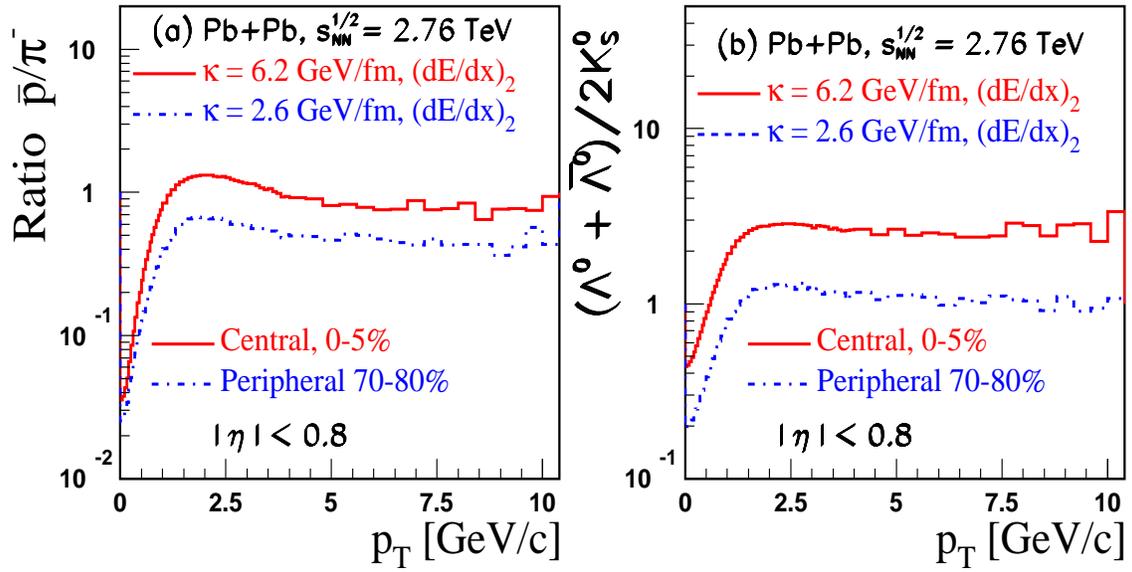}
\vskip 0.5cm\caption[Centrality dependence of (non)strange particle ratios]
{\small (Color online) 
Predictions for centrality dependence  
of (non)strange baryon/meson ratios from {\small HIJING/B\=B v2.0}.
The results include SCF effects and J\=J loops.
In each figure the histograms correspond to two value of centrality 
and the associated string tension.
\label{fig:fig6}
}
\end{figure}

With the best set of parameters used to describe charged particle production
in central (0-5\%)  and peripheral (70-80\%) 
Pb+Pb collisions at $\sqrt{s_{\rm NN}}$ = 2.76 TeV (see Sec.~IIIA),
we present in Fig.~\ref{fig:fig6} the model predictions
for centrality dependence of
the nonstrange baryon/meson ratio (left panel) and of the 
strange baryon/meson ratio (right panel).
Our model predicts that these ratios do not depend strongly 
on energy and centrality. The model also predicts that the baryon/meson anomaly
will persist up to high $p_T$.
For central (0-5\%) Pb+Pb collisions at $\sqrt{s_{\rm NN}}$ = 2.76
TeV, in a scenario with jet quenching and shadowing as explained in
Sec.~II, but without J\=J loops and SCF effects 
the $\bar{p}/\pi^-$ ratio in the region of the  
maximum ($2 < p_T < 3$ GeV/{\it c}) is 0.07. 
We have studied the sensitivity of the results 
to the J\=J loops and SCF effects.
Considering only J\=J loops leads to an increase by a factor of
two, and taking into consideration only SCF effects 
leads to an increase up to a factor of  $\approx$ 14 
for the $\bar{p}/\pi^-$ ratio.
Note that an important fraction of this increase is also due to
much stronger quench of pions in comparison with that of antiprotons, 
as we will show below (see Fig.~\ref{fig:fig8}).


The ALICE and CMS detectors are designed to perform measurements in the 
high-multiplicity environment of central Pb+Pb collisions at the LHC 
\cite{Aamodt:2011zj}, \cite{Khachatryan:2011tm}.
To isolate QGP signatures in heavy-ion collisions, 
understanding the particle production properties 
from $NN$ interactions is necessary. 
In addition, the transverse momentum spectrum in $p+p$ collisions 
serves as a baseline to study possible suppression 
(i.e., nuclear modification factor $R_{\rm AA}^{\rm ID}(p_T)$)
in heavy-ion collisions. This stresses the need for reference $p+p$ 
measurements at LHC energies of identified particles.
In a previous paper \cite{ToporPop:2010qz} we show that the large 
(strange)baryon-to-meson ratios measured at Tevatron (1.8 and 1.96 TeV) 
and LHC energies (0.9 and 7 TeV) are
well described in the framework of the {\small HIJING/B\=B} v2.0 model.
Recently, the CDF reported a set of measurements of inclusive invariant 
$p_T$ differential cross sections of hyperons with different strangeness,
$\Lambda^0$ (quark content ({\it uds})), $\Xi^-$ ({\it dss}), and $\Omega$ 
({\it sss}). These data are obtained for the central region with pseudorapidity
range $-1<\eta<1$ and $p_T$ up to 10 GeV/{\it c}, and are used to test 
our model calculations which include SCF effects and J\=J loops.

In a scenario with SCF effects (i.e., considering 
$\kappa=\kappa(s)$ \cite{ToporPop:2010qz}) the increase of the yield 
is higher for multistrange hyperons ($\Xi^-$,$\Omega^-$) 
than for the strange hyperon ( $\Lambda^0$),
in comparison with the results obtained without SCF effects 
(i.e., taking vacuum string tension value $\kappa_0 = 1 $ GeV/fm).
The increase of the yield with strangeness content is due to an increase 
of strange quark suppression ($\gamma_s$), 
obtained from a power-law dependence of effective string tension values,
$\kappa(s) = \kappa_{0} \,\,(s/s_{0})^{0.06}\,\,{\rm GeV/fm}$
(see Fig.~2 from Ref.~\cite{ToporPop:2010qz}).

\begin{figure}[h!]

\centering

\includegraphics[width=0.9\linewidth,height=7.0cm]{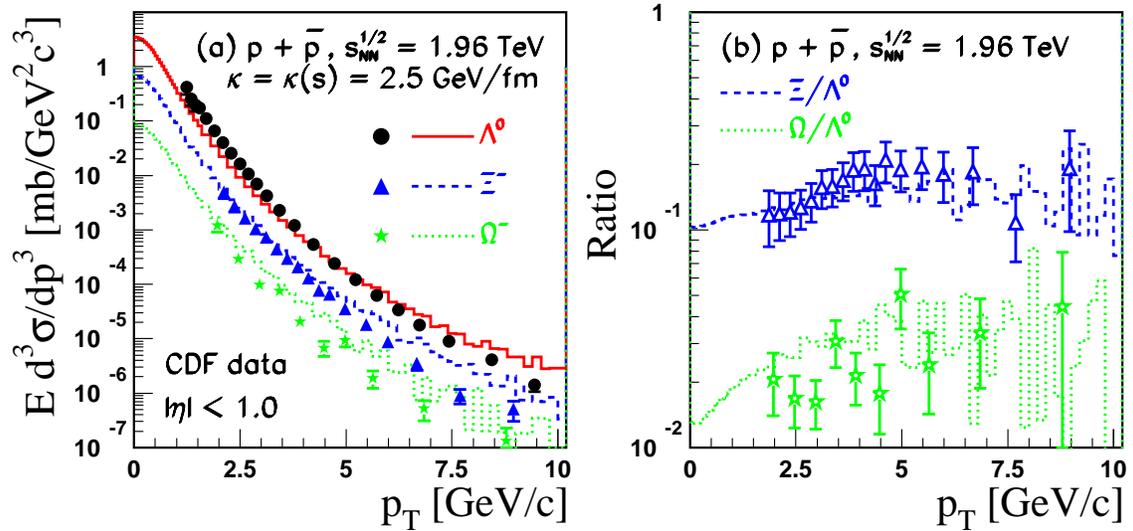}
\vskip 0.5cm\caption[strange prod pp at 1.96 TeV]
{\small (Color online) (a) Comparison of {\small HIJING/B\=B v2.0} 
predictions for the $P_T$ differential cross sections of $\Lambda^0$ (solid),
$\Xi$ (dashed), and $\Omega$ (dotted histograms). The calculations are for 
$p+\bar{p}$ collisions at 1.96 TeV and include SCF effects and J\=J loops 
as in Ref. \cite{ToporPop:2010qz}. 
(b) The ratios $\Xi/\Lambda^0$ (dashed histogram) and
$\Omega/\Lambda^0$ (dotted histogram).
The data are from the CDF Collaboration 
\cite{Aaltonen:2011wz}. Only statistical uncertainties are shown.
\label{fig:fig7}
}

\end{figure}

Figure~\ref{fig:fig7} shows the predicted $p_T$ differential cross 
sections (left panel) for the three hyperon resonances in comparison with data.
The plots on the right side of Fig.~\ref{fig:fig7} show the ratio of the 
$p_T$ differential cross sections for $\Xi^-/\Lambda^0$
and $\Omega^-/\Lambda^0$. In the limit of the experimental error  
(systematics+statistical) our model describes well these data.

Strangeness enhancement, strong baryon transport, and increase of 
intrinsic transverse momentum are all expected consequences of 
longitudinal SCF effects. These are modeled in our microscopic models 
as an increase of the effective string tension that controls 
quark-antiquark ($q\bar{q}$) and diquark-antidiquark (qq$\bar{\rm qq}$) 
pair creation rates and the strangeness suppression factors ($\gamma_s$).
The predictions in central Pb+Pb collisions  
for initial energy density and temperature are 
$\epsilon_{\rm LHC} \approx 200$ GeV/fm$^3$ and $T_{\rm LHC} \approx 500$
MeV, respectively \cite{Muller:2006ru}. 
Both values would lead in the framework of our model
to an estimated increase of the average value of string tension to 
$\kappa_{\rm LHC} \approx 5-6$ GeV/fm \cite{armesto2_08}, 
consistent with our parametrization from Eq.~(5).

\begin{figure} [h!]

\centering

\includegraphics[width=0.9\linewidth,height=7.5cm]{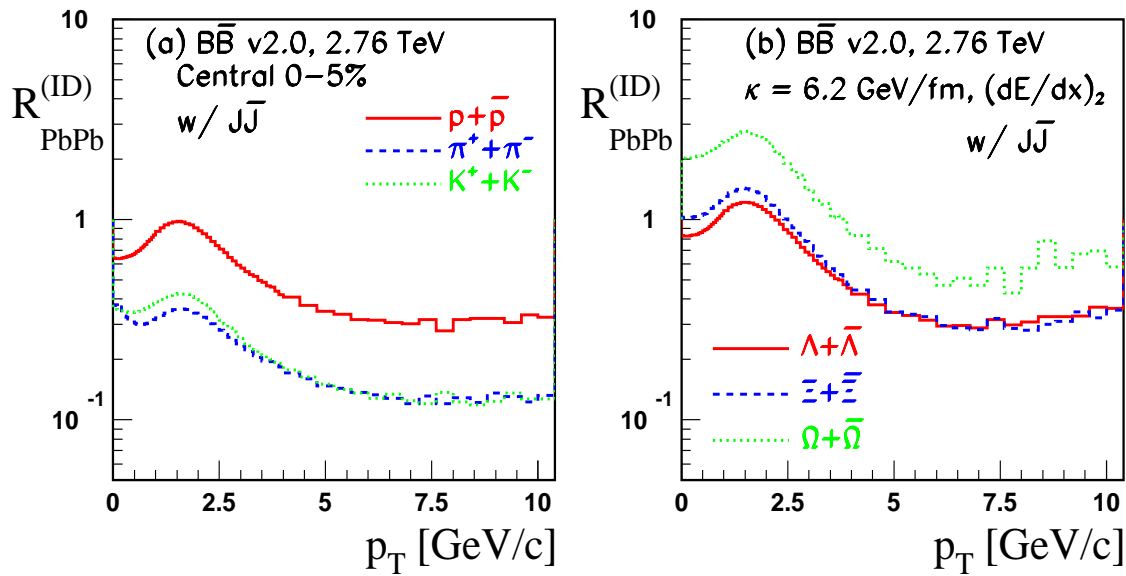}
\vskip 0.5cm\caption[$R_{AA}$ for ID particles-central Pb-Pb 0-5\%]
{\small (Color online) 
Predictions from {\small HIJING/B\=B v2.0} model for
nuclear modification factors $R_{\rm AA}^{\rm ID}$ of identified particles
at mid-pseudorapidity ($-0.8<\eta<0.8$) in central (0-5\%) Pb+Pb collisions at 
$\sqrt{s_{\rm NN}}$ = 2.76 TeV. The results include SCF effects and J\=J loops.
Left: pions, kaons and protons. Right: hyperons $\Lambda, \Xi, \Omega$.
\label{fig:fig8}
}
\end{figure}

The predictions for NMFs of identified particles ($R_{\rm AA}^{\rm ID}(p_T)$)
are presented in Fig.~\ref{fig:fig8} for nonstrange particles (left panel)
and (multi)strange particles (right panel).
From the results of our model we conclude that the baryon/meson anomaly,
i.e., different meson and baryon suppression, 
will persist in central Pb+Pb collisions at the LHC energy
$\sqrt{s_{\rm NN}}$ = 2.76 TeV up to a higher value of transverse momentum.
The $R_{\rm AA}^{\rm ID}(p_T)$ also exhibit an 
ordering with strangeness content at low and intermediate $p_T$, the 
increase of the yield being higher for multistrange hyperons than
that for (non)strange baryons 
(i.e., $R_{\rm PbPb}(\Omega) >R_{\rm PbPb}(\Xi) >  R_{\rm PbPb}(\Lambda)> 
 R_{\rm PbPb}(p)$).
This prediction could be verified by the future 
planned experiments at the LHC.

\section{Summary and Conclusions}

We have studied the influence of strong longitudinal color fields
and of possible multigluon J\=J loops dynamics 
in particle production in Pb+Pb  collisions at 2.76 TeV.
We have investigated a set of observables
sensitive to the dynamics of the collisions,
covering both longitudinal and transverse degree of freedom.
A detailed comparison with 
available experimental data from the LHC has been performed.

We find that the inclusion of the multiple minijet source 
limits the growth of the string tension $\kappa(s)$ to be approximately 
linear as a function of saturation scale $Q_{\rm sat}$, in contrast to recent
approaches \cite{McLerran:2010ex} where $\kappa(s)$ scales as $Q_{\rm sat}^2$.
The reason is that in the CGC model the collinear factorized minijet 
mechanism is suppressed by geometric scaling to much higher $p_T$. 
Moreover, an increase with $A$ due to an increase in initial energy density 
is necessary in order to better describe new ALICE data.   

We have shown that strong color field (SCF) could play an important role in
particle production at mid-rapidity in heavy-ion collisions.
The mechanisms of hadron production
are very sensitive to the early phase of the
collisions, when fluctuations of the color field strength are highest.
Strong color field effects are modeled
by varying the effective  string tension that controls the
quark-antiquark ($q\bar{q}$) and 
diquark-antidiquark (${\rm qq}\overline{\rm qq}$) pair creation rates
and strangeness suppression factors ($\gamma_s$).
SCF also modify the fragmentation processes 
resulting in an increase of (strange)baryons.

We show that the baryon/meson anomaly manifests in 
an increase of baryon-to-meson ratios
as well as in different suppression of mesons and baryons and persists up
to high $p_T$ ( $p_T < 10$ GeV/{\it c}). 
 We show also that the $R_{\rm AA}^{\rm ID}(p_T)$ exhibit an 
ordering with strangeness content at low and intermediate $p_T$, the 
increase of the yield being higher for multistrange hyperons than
that for (non)strange baryons.

\section{Acknowledgments}
\vskip 0.2cm 

{\bf Acknowledgments:} We thank S. Das Gupta, S. Jeon, and P. Levai 
for fruitful discussions and support.  
VTP acknowledges the use of computer facilities at Columbia
University, New York, where part of these calculations were performed.
This work was supported by the Natural Sciences and Engineering 
Research Council of Canada.  
This work was also supported by the Division of Nuclear Science, 
of the U.S. Department of Energy, under Contract No. DE-AC03-76SF00098 and
DE-FG02-93ER-40764.

\end{document}